**Zwanzig-Mori projection operators and EEG dynamics: deriving a simple equation of motion**

David Hsu and Murielle Hsu

Department of Neurology, University of Wisconsin, Madison WI, USA

**Corresponding author:** David Hsu

**Email addresses:**
DH: hsu@neurology.wisc.edu
MH: murielle.hsu@gmail.com

**Abstract**

We present a macroscopic theory of electroencephalogram (EEG) dynamics based on the laws of motion that govern atomic and molecular motion. The theory is an application of Zwanzig-Mori projection operators. The result is a simple equation of motion that has the form of a generalized Langevin equation (GLE), which requires knowledge only of macroscopic properties. The macroscopic properties can be extracted from experimental data by one of two possible variational principles. These variational principles are our principal contribution to the formalism. Potential applications are discussed, including applications to the theory of critical phenomena in the brain, Granger causality and Kalman filters.

# 1. Introduction

The electrical activity of the brain has intrigued scientists since the invention of the electroencephalogram (EEG) [1, 2]. Scalp and intracranial EEG's are now in widespread clinical use in the diagnosis and management of epilepsy and other neurological disorders. These applications rely on empirical correlations between certain EEG patterns with specific neurological disorders. Intense interest exists in trying to understand EEG dynamics at a deeper level, so as to extract ever more information about brain health and function. These efforts fall into two classes: those which are largely empirical, based on traditional correlations between EEG patterns and clinical observations, and those which are theory-based, where one has in mind a certain model of brain dynamics and then one tries to interpret EEG patterns in terms of the theoretical model. In the empirical class are recent efforts to correlate high frequency oscillations with epileptogenic tissue [3]. In the theory-based class, the most celebrated approach is the cable theory of Hodgkin and Huxley [4]. This theory can be scaled up using compartmental models to describe networks of thousands or even millions of neurons using high power computers. In the hands of a master, much insight can come from such simulations [5]. However, these methods are computationally intensive. They are not easily scaled to truly macroscopic levels and they are not amenable to the clinical diagnostic situation where one wants to know, for specific EEG samples from specific individuals, whether a certain brain pathology is present.

Mesoscopic and macroscopic level theories of EEG dynamics have also been proposed, each based on a plausible basic postulate or mathematical model of the electrical activity of the brain [2, 6-8]. The methods of nonlinear dynamics (chaos theory) also fall into this class and have been applied to seizure prediction [9, 10].

It would be desirable to base a macroscopic theory of EEG dynamics on fundamental physical principles, for instance, on the laws of motion that govern atomic and molecular motion. In this paper, we discuss such a theory. The approach we take was developed by Zwanzig [11] and Mori [12] to explain thermodynamic irreversibility (why entropy always increases). The result is a simple equation of motion that has the form of a generalized Langevin equation (GLE), which requires knowledge only of macroscopic properties. The macroscopic properties can be extracted from experimental data by one of two variational principles. Potential applications are discussed, including applications to epilepsy research, the theory of critical phenomena in brains, Granger causality and Kalman filters.

**2. Theory**

EEG dynamics results from the motion of microscopic charged particles in the brain. The charged particles interact with other charged particles as well as with uncharged particles. The total number of particles is astronomical, on the order of $10^{23}$. A macroscopic EEG electrode either on the scalp or inserted into the brain detects voltages set up by the local distribution of the charged particles. The changes in these voltages reflect the local flux of charged particles. The forces exerted by the particles on each other are highly nonlinear. How can we possibly hope to derive an equation of motion for the macroscopic EEG voltages? The most elegant solution is to apply Zwanzig-Mori projection operators [11, 12]. We follow the discussion of Zwanzig [13]; see also Berne and Pecora [14] for more detailed discussions.

Let the voltage measured by electrode $n$ at time $t$ be represented as $x(n,t)$ where $n$ = 1 to $N$. If the reference ground is taken to be infinitely far away, then $x(n,t)$ is given by:

$$x(n,t) = \sum_{i=1}^{N_0} \frac{Q_i}{|\vec{r}_i(t) - \vec{r}_n|}. \qquad \text{Eq (1)}$$

where $N_0$ is the total number of charged particles, $Q_i$ is the charge of particle $i$, $\vec{r}_i(t)$ is its position at time t, and $\vec{r}_n$ is the position of the $n^{th}$ electrode. If the reference ground is chosen differently, then $x(n,t)$ will be given by some linear combination of the terms on the right hand side of Eq (1). The choice of reference ground does not affect what follows. Next, let $x(n,t)$ represent the $n^{th}$ component of an $N$-dimensional vector $X(t)$. Hence, the voltage readouts from an $N$-channel EEG are the $N$ components of the vector $X(t)$. For convenience, we will take $X(t)$ to represent the EEG voltage after the time average of the raw data has been subtracted out: $X(t) = X_{raw}(t) - \langle X_{raw}(t) \rangle$.

In what follows, we will refer to the $x(n,t)$'s as the *explicit* degrees of freedom, while the $\vec{r}_i(t)$'s and all other degrees of freedom are the *implicit* or "bath" degrees of freedom. At this point, readers who wish to skip some of the more mathematical discussion may jump to Eq. (17).

To continue, if the microscopic particles of the system all obey the laws of physics, then the dynamics of $X(t)$ can be written:

$$\frac{\partial}{\partial t} X(t) \equiv \dot{X}(t) = LX(t) \qquad \text{Eq (2)}$$

Here $L$ is the Liouville operator. The classical and quantum Liouville operators, respectively, are defined as:

$$L = \sum_{i=1}^{N_0} \left\{ \frac{\partial H}{\partial \vec{p}_i} \cdot \frac{\partial}{\partial \vec{q}_i} - \frac{\partial H}{\partial \vec{q}_i} \cdot \frac{\partial}{\partial \vec{p}_i} \right\} \qquad \text{Eq (3a)}$$

$$L = \frac{i}{\hbar}[H, \ ] \qquad \text{Eq (3b)}$$

Here $\vec{q}_i$ and $\vec{p}_i$ are the coordinate and momentum of the $i^{th}$ particle, $H$ is the total Hamiltonian, $\hbar$ is Planck's constant, and the square brackets denote the quantum commutator. In what follows, one may choose the dynamics to be determined by either the classical or quantum Liouville operator.

The Hamiltonian $H$ in Eq (3) contains kinetic and potential energy terms for all the microscopic particles in the system. The potential energy terms describe how every microscopic particle interacts with every other microscopic particle. These terms are in general highly nonlinear. We shall not need to know the details of these interactions. The Zwanzig-Mori approach only requires that the microscopic dynamics obeys the form of Eq (2). The realm of validity for what follows thus includes all of classical and quantum mechanics.

Equations (2)-(3) are formally correct but they are not useful for describing macroscopic systems in practice because we cannot possibly specify $L$ for all $N_0 \approx 10^{23}$ microscopic particles. If we could formally "average" or integrate Eq (2) over all the microscopic degrees of freedom $\bar{q}_i$ and $\bar{p}_i$ (that is, over all the *implicit* degrees of freedom) without integrating out the *explicit* degrees of freedom contained in $X(t)$, then we can transform Eq (2) into a useful equation of motion for $X(t)$. Here is where projection operators come in. Let $\Gamma$ represent all the microscopic particle coordinates and momenta (i.e., it represents a point in phase space), and let $d\Gamma = d\bar{q}_1 \, d\bar{p}_1 \, d\bar{q}_2 \, d\bar{p}_2 ... d\bar{q}_{N_0} \, d\bar{p}_{N_0}$. Define an inner product

$$(A, B) = \int d\Gamma \, \rho(\Gamma) \, A(\Gamma) \, B^*(\Gamma) \qquad \text{Eq (4)}$$

where $\rho(\Gamma)$ gives a microscopic distribution of $\Gamma$, *not necessarily at equilibrium*, and where the asterisk denotes a complex conjugate. Here $A$ and $B$ are explicit or implicit degrees of freedom which for the time being we will allow to be complex (i.e., they may have both real and imaginary components). We will later show how to recover convenient expressions for purely real observables, such as the EEG voltages and their time derivatives. Note that Eq (4) is essentially a phase space average of the product of $A$ times $B^*$.

Now imagine that we identify the vector $A$ as the explicit degree of freedom, and we wish to project or integrate out all other degrees of freedom from Eq (2). The projection operator with respect to $A$ is:

$$P = (..., A) \, (A, A)^{-1} A \qquad \text{Eq (5)}$$

This operator integrates out all degrees of freedom of the system except those that affect $A$. Applying this projection operator on Eq (2) will result in an equation of motion for $A$ only. We will not reproduce here the mathematical manipulations described in detail in Refs [12-14]. One needs the operator identity:

$$e^{(t-t_0)L} = e^{(t-t_0)(1-P)L} + \int_0^{t-t_0} d\tau \, e^{(t-\tau)L} \, PLe^{\tau(1-P)L} \qquad \text{Eq (6)}$$

We will take $t_0$ to represent the initial time within an arbitrary time window. After some work, for times such that $t \geq t_0$, one arrives at the generalized Langevin equation (GLE):

$$\dot{A}(t) = i\Omega A(t) - \int_0^{t-t_0} d\tau \, \Gamma(\tau) A(t-\tau) + F(t) \qquad \text{Eq (7)}$$

where

$$i\Omega = (LA, A)(A, A)^{-1} \qquad \text{Eq (8)}$$

$$F(t) = e^{(t-t_0)(1-P)L}(1-P)LA \qquad \text{Eq (9)}$$

$$\Gamma(t) = -(LF(t), A)(A, A)^{-1} \qquad \text{Eq (10)}$$

Here $\Omega$ has the interpretation of a frequency, $F(t)$ is a "random force" due to the dynamics of the implicit degrees of freedom, and $\Gamma(\tau)$ is referred to as a memory function that defines the effect of prior values of $A$ on its current time rate of change. Equation (7) is the desired equation of motion for the explicit degrees of freedom. An important formal result is that

$$(F(t), A(t_0)) = 0 \qquad \text{Eq (11)}$$

which states that the random force acting on the macroscopic observable *A* is *not correlated* with *A*. This result is rigorously true and it underlies the foundation of the Onsager regression hypothesis [14]. It will also suggest one of the variational principles by which we extract model parameters from experiment, as we will discuss shortly.

Other forms of the GLE may be derived from Eq (7). For convenience, let us rewrite Eq (7) as

$$\dot{A}(t) = -W \circ A(t) + F(t) \qquad \text{Eq (12)}$$

where the open circle denotes a convolution and the terms in $\Omega$ and $\Gamma(\tau)$ have both been absorbed into the convolution:

$$W \circ A(t) \equiv -i\Omega A(t) + \int_0^{t-t_0} d\tau\, \Gamma(\tau)\, A(t-\tau) \qquad \text{Eq (13)}$$

The macroscopic observable *A*(*t*) thus far is allowed to have both real and imaginary parts. Now consider a purely real vector *A*(*t*) of the form:

$$A(t) = \begin{bmatrix} X(t) \\ V(t) \end{bmatrix} \qquad \text{Eq (14)}$$

We choose this form because the underlying dynamics is governed by the laws of physics, for which the coordinate and momentum (here represented as the velocity) are independent degrees of freedom. The coordinate *X*(*t*) is constrained such that

$$\frac{d}{dt} X(t) = V(t) \equiv \dot{X}(t) \qquad \text{Eq (15)}$$

In practice, recall that *X*(*t*) represents the *N* channels of EEG voltages. Hence, *V*(*t*) is the first time derivative of *X*(*t*), which in practice can be constructed from experimental time series data by taking finite differences, e.g., $V(t) \approx [X(t+\delta t) - X(t-\delta t)]/(2\delta t)$, where $\delta t$ is the EEG sampling time.

Taking the formal time derivative of Eq (14) and inserting Eq (12), we find

$$\begin{bmatrix} \dot{X}(t) \\ \dot{V}(t) \end{bmatrix} = -\begin{bmatrix} W_{11} & W_{12} \\ W_{21} & W_{22} \end{bmatrix} \circ \begin{bmatrix} X(t) \\ V(t) \end{bmatrix} + \begin{bmatrix} F_1(t) \\ F_2(t) \end{bmatrix} \qquad \text{Eq (16)}$$

Applying the constraint of Eq (15), we have

$$\ddot{X}(t) = -K \circ X(t) - G \circ \dot{X}(t) + F_R(t) \qquad \text{Eq (17a)}$$

where $\ddot{X}(t) = \dot{V}(t)$, $K = W_{21}$, $G = W_{22}$ and $F_R(t) = F_2(t)$. This equation may be written in more expanded form as

$$\ddot{X}(t) = -\int_0^{t-t_0} d\tau\, K(\tau) X(t-\tau) - \int_0^{t-t_0} d\tau\, G(\tau) \dot{X}(t-\tau) + F_R(t) \qquad \text{Eq (17b)}$$

Equation (17b) has the form of an equation of motion for a time-delayed damped harmonic oscillator with force constant matrix $K$, friction constant matrix $G$ and random force $F_R(t)$. The purely real matrix $K$ has convolution components $K(m,n,\tau)$ giving the influence of $x(n, t-\tau)$ on $x(m,t)$ with time delay $\tau$. That is, a fluctuation in the $n^{th}$ macroscopic coordinate of magnitude $x(n)$ at a certain point in time, results in a force on the $n^{th}$ macroscopic coordinate of magnitude $-K(m,n,\tau)x_n$ at a time $\tau$ later. Similarly, $G(\tau)$ is a purely real time-delayed friction matrix with components $G(m,n,\tau)$ giving the frictional force of $\dot{x}(n, t-\tau)$ on the $m^{th}$ macroscopic coordinate with time delay $\tau$.

Recall that $X(t)$ represents the EEG voltages. Hence Eq (17) tells us that we can think of EEG dynamics as being driven by three kinds of forces: one due to interactions between neurons at the sites of the electrodes (the $K$-term), another due to frictional forces (the $G$-term), and a third due to random environmental noise (the $F_R(t)$-term).

The influence of the enormous number of implicit degrees of freedom are hidden in microscopic expressions for the $K(\tau)$, $G(\tau)$ and $F_R(t)$ terms as expressed through Eqs (8)-(10). In certain highly idealized situations, one may be able to estimate these functions quantitatively. In the general case, however, Eqs (8)-(10) are computationally intractable. Nonetheless, the functions $K(\tau)$, $G(\tau)$ and $F_R(t)$ also represent *macroscopic properties of the system*, and it should be possible to extract them from experiment. What we need is a way to relate these functions to experimental data, preferably by a variational principle.

There are two obvious approaches to a variational principle by which $K(\tau)$, $G(\tau)$ and $F_R(t)$ may be extracted from experiment: (*I*) minimization of the amplitude of the random force $F_R(t)$ and (*II*) minimization of correlation of the random force with the macroscopic coordinate.

*2.1. Variational principle I: minimization of random force amplitude*

Equation (17) represents a way of dividing up the total "force" acting on the macroscopic observables into three components: (1) that due to explicit interactions between the macroscopic observables, represented by the term in $K$, (2) that due to frictional dissipation, represented by the term in $G$ and (3) that due to environmental fluctuations, represented by the term in $F_R(t)$. One may plausibly argue that one should ascribe as much of the total force as possible to the explicit interactions between the macroscopic observables, and as little as possible to environmental fluctuations. In this case, one may define an error functional which adds up all the square amplitudes of $F_R(t)$ over the entire time interval under observation, $t = [t_a, t_b]$, with the goal of minimizing it:

$$E_I(t_a, t_b) = \tfrac{1}{2} \int_{t_a}^{t_b} dt \ F_R(t) \cdot F_R(t). \qquad \text{Eq (18)}$$

To minimize this error functional, one may take each value of $K(m,n,\tau)$ and $G(m,n,\tau)$, for each $m$, $n$ and $\tau$, as an independent parameter, and vary them so as to make $E_I(t_a, t_b)$ as small as possible. To do this, first insert Eq (17) into Eq (18), to express $E_I(t_a, t_b)$ in terms of $X(t)$, $\dot{X}(t)$, $\ddot{X}(t)$, $K(\tau)$, and $G(\tau)$. Next, insert experimental values for $X(t)$, $\dot{X}(t)$ and $\ddot{X}(t)$ into $E_I(t_a, t_b)$. One may then minimize $E_I(t_a, t_b)$ with respect to $K(\tau)$ and $G(\tau)$ by standard algorithms [15]. Because $E_I(t_a, t_b)$ is quadratic in $K(\tau)$ and $G(\tau)$, one is guaranteed a unique solution. This procedure represents our first variational principle.

For future reference, we list the derivatives of $E_I(t_a, t_b)$ with respect to the $K(m,n,\tau)$'s and $G(m,n,\tau)$'s. These equations are useful in global minimization algorithms [15]:

$$\frac{\partial E_I(t_a,t_b)}{\partial K(m,n,t')} = \int_{t'+t_a}^{t_b} dt \left[ \ddot{x}(m,t) + \sum_{k=1}^{N} K_{m,k} \circ x(k,t) + \sum_{k=1}^{N} G_{m,k} \circ x(m,t) \right] x(n, t-t')$$

Eq (19)

$$\frac{\partial E_I(t_a,t_b)}{\partial G(m,n,t')} = \int_{t'+t_a}^{t_b} dt \left[ \ddot{x}(m,t) + \sum_{k=1}^{N} K_{m,k} \circ x(k,t) + \sum_{k=1}^{N} G_{m,k} \circ x(m,t) \right] \dot{x}(n, t-t')$$

Eq (20)

Alternatively, after some experience, if one finds that $K(m,n,\tau)$ and $G(m,n,\tau)$ tend to decay exponentially, possibly with some oscillations, then one may try to save some computational effort by parameterizing $K(m,n,\tau)$ and $G(m,n,\tau)$, for example, using

$$K(m,n,\tau) = A_K(m,n) \exp[-b_K(m,n)\,\tau] \cos[\omega_K(m,n)\tau] \qquad \text{Eq (21)}$$

$$G(m,n,\tau) = A_G(m,n) \exp[-b_G(m,n)\,\tau] \cos[\omega_G(m,n)\tau] \qquad \text{Eq (22)}$$

The minimization of $E_I(t_a, t_b)$ would then be with respect to $A_K(m,n)$, $b_K(m,n)$, $\omega_K(m,n)$, etc. The drawback here is that the minimization algorithm would require an iterative algorithm suitable for nonlinear fits, and a unique solution is not guaranteed [15].

*2.2. Variational principle II: minimization of random force correlation*

To formulate our second variational principle, consider Eq (11), which states that the random force should not be correlated with any macroscopic observable when averaged over the phase space available to the microscopic degrees of freedom, as defined in Eq (4). One may hypothesize that this phase space average may be replaced by an average *over time*. This hypothesis is known as the *ergodic hypothesis*. Note that in Eq (4), we do not assume the phase space average is necessarily an equilibrium average. If for some possibly non-equilibrium distribution function, the ergodic hypothesis holds true, then we may take the assumption that the bath random force is not time-correlated with macroscopic observables as the basis for our second variational principle. There are subtleties in this approach which must be treated with care.

It is generally not possible to be certain whether an experimental system is ergodic or not. For instance, the explicit degrees of freedom may happen to be trapped in a portion of phase space that is separated from other regions of phase space by very high free energy barriers. Even if over time the explicit degrees of freedom sample all of the phase space available within this bounded region, we can never know that there are not other regions that would have been equally sampled, were it not for the impenetrable energy barriers separating the regions. In this case, we can still *define* the random force term in Eq (17) to be such that it is not time-correlated with the macroscopic observables. If the random force were correlated with a macroscopic observable, then of course it would not really be random.

In what follows, we will assume that the bath degrees of freedom are ergodic. We will be careful, however, not to invoke the ergodic hypothesis for correlations between two explicit degrees of freedom, because we wish to allow for the possibility that the explicit degrees of freedom are non-ergodic and far from equilibrium.

To proceed, let us define a time averaged correlation function with the time average being over a time period $[t_a, t_b]$, with $t_a \leq t_0 \leq t_b$:

$$C(n, m, t) = \int_{t_a}^{t_b} d\tau \, x(n, t+\tau) \, x(m, \tau) \equiv \langle x(n,t) \, x(m,0) \rangle \qquad \text{Eq (23)}$$

Let $C(n,m,t)$ represent the $(n,m)$ matrix element of a matrix $C(t)$. Evaluate Eq (17b) at time $t+t_0$, multiply both sides by $X(t_0)$, and integrate over $t_0$ over the range $[t_a, t_b]$. We will assume that $K(\tau)$, $G(\tau)$ and $F_R(t)$ do not depend on the time $t_0$, i.e., we assume these functions are *stationary*. The result is:

$$\ddot{C}(t) = -K \circ C(t) - G \circ \dot{C}(t) + C_{RX}(t) \qquad \text{Eq (24a)}$$

which may be written in more expanded form as

$$\ddot{C}(t) = -\int_0^t d\tau \, K(\tau) C(t-\tau) - \int_0^t d\tau \, G(\tau) \dot{C}(t-\tau) + C_{RX}(t) \qquad \text{Eq (24b)}$$

where

$$C_{RX}(t) = \langle F_R(t) X(0) \rangle. \qquad \text{Eq (25)}$$

Here $C_{RX}(t)$ is the time correlation between the random force and the macroscopic observables, when averaged over time. If we take the bath random force to be ergodic, then the time averaged correlation $C_{RX}(t)$ from Eq (25) is equal to the phase space averaged correlation $(F(t), X(t_0))$ from Eq (11), and therefore $C_{RX}(t)$ should be equal to zero. Minimizing the square amplitude of every element of the matrix $C_{RX}(t)$ over the time interval $[t_a, t_b]$ represents our second variational principle.

The careful reader will note that Eq (24) is not quite the same as the typical equation of motion for the time correlation function, as defined in standard texts [13, 14]. In the standard derivations, an equation of motion that is identical in form as Eq (24) is derived but the time correlation functions involve an average over phase space, not over time. One then invokes the ergodic hypothesis to equate these phase space averaged correlation functions with time averaged correlation functions. However, we wish to avoid making the ergodic hypothesis for the explicit degrees of freedom. In our derivation, we invoke the ergodic hypothesis only for the *bath* degrees of freedom.

To continue, define an error functional that consists of the sum of the square of every element of the matrix $C_{RX}(t)$, summed over all the elements of the matrix and over every instant in time in the time interval $[t_a, t_b]$:

$$E_{II}(t_a, t_b) = \frac{1}{2} \int_{t_a}^{t_b} dt \, Tr\left[ C_{RX}(t) \, C_{RX}(t)^T \right] \qquad \text{Eq (26)}$$

Here *Tr* signifies a matrix trace and *T* a matrix transpose. To minimize the error functional of Eq (26), one may take each value of $K(m,n,\tau)$ and $G(m,n,\tau)$, for each *m, n* and $\tau$, as an independent parameter, and vary them so as to make $E_{II}(t_a, t_b)$ as small as possible. To do this, first insert Eq (24) into Eq (26), to express $E_{II}(t_a, t_b)$ in terms of $C(t)$, $\dot{C}(t)$, $\ddot{C}(t)$, $K(\tau)$, and $G(\tau)$. Next, insert experimental values for $C(t)$, $\dot{C}(t)$, $\ddot{C}(t)$ into $E_{II}(t_a, t_b)$. One may then minimize $E_{II}(t_a, t_b)$ with respect to $K(\tau)$ and $G(\tau)$ by standard algorithms [15]. Because $E_{II}(t_a, t_b)$ is quadratic in $K(\tau)$ and $G(\tau)$, one is guaranteed a unique solution. This procedure represents our second variational principle.

For future reference, we also list the derivatives of $E_{II}(t_a, t_b)$ with respect to the $K(m,n,\tau)$'s and $G(m,n,\tau)$'s. These equations are useful in global minimization algorithms [15]:

$$\frac{\partial E_{II}(t_a,t_b)}{\partial K(m,n,t')} = \int_{t'+t_a}^{t_b} dt \left[ \ddot{C}(t) C^T(t-t') + [K \circ C(t)] C^T(t-t') + [G \circ \dot{C}(t)] C^T(t-t') \right]_{m,n}$$

Eq (27)

$$\frac{\partial E_{II}(t_a,t_b)}{\partial G(m,n,t')} = \int_{t'+t_a}^{t_b} dt \left[ \ddot{C}(t) \dot{C}^T(t-t') + [K \circ C(t)] \dot{C}^T(t-t') + [G \circ \dot{C}(t)] \dot{C}^T(t-t') \right]_{m,n}$$

Eq (28)

Alternatively, after some experience, if one finds that $K(m,n,\tau)$ and $G(m,n,\tau)$ tend to decay exponentially, possibly with some oscillations, then one may try to save some computational effort by parameterizing $K(m,n,\tau)$ and $G(m,n,\tau)$, for example, using

$$K(m,n,\tau) = A_K(m,n) \exp[-b_K(m,n)\tau] \cos[\omega_K(m,n)\tau] \qquad \text{Eq (29)}$$

$$G(m,n,\tau) = A_G(m,n) \exp[-b_G(m,n)\tau] \cos[\omega_G(m,n)\tau] \qquad \text{Eq (30)}$$

The minimization of $E_{II}(t_a, t_b)$ would then be with respect to $A_K(m,n), b_K(m,n), \omega_K(m,n)$, etc. The drawback here is that the minimization algorithm would require an iterative algorithm suitable for nonlinear fits, and a unique solution is not guaranteed [15].

**3. Discussion**

The mathematical or physical model that one chooses to represent EEG dynamics influences how one interprets experimental observations. The model itself acts as a filter which biases one's interpretations. It is desirable therefore that these models be based on fundamental physical principles. The model we present here is suitable for interpreting electrophysiological data acquired from extracellular recordings. It is a macroscopic level model which complements more microscopic Hodgkin-Huxley type models. The assumptions that we make are fourfold. First, we assume that the underlying dynamics obeys the laws of either classical or quantum physics. This assumption is quite safe for biological systems.

Second, we assume that the total system under consideration is isolated and free from external influences, which allows us to write the Liouville operator in Eq (2) as being free of an explicit time dependence, i.e., the overall system is *stationary*. If there is an external driving force, such as environmental stimuli driving a learning experience, then the Liouville operator will have an explicit time dependence and the operator identity of Eq (6) is no longer valid. On the other hand, one could always subsume the external degrees of freedom into the Liouville operator, to make the external degrees of freedom part of the total system under consideration. It does not matter how many degrees of freedom are subsumed into the Liouville operator in this way, since we integrate them out anyway. One could in principle claim to include the entire universe in one's Liouville operator, in which case, neglecting relativistic and other cosmological effects, one recovers the time independence of the Liouville operator and the derivation of Eq (17) proceeds as above.

Thirdly, we have made the ergodic assumption for the bath degrees of freedom, which assumes that the bath degrees of freedom can quickly (on time scales much faster than the time scales of the explicit degrees of freedom) explore all of the phase space available to the bath. Importantly, we have avoided making the ergodic hypothesis for the explicit degrees of freedom, and thus our principal theoretical result, Eq (17), is capable of describing systems far from equilibrium.

The fourth assumption is comprised of either variational principle *I* or *II*, either of which allows us to extract the model parameters from experimental data. Variational principle *I* tries to explain as much of the dynamics as possible using the macroscopic observables, ascribing as little as possible to the influence of random unseen forces. Variational principle *II* does not assume that the random forces are small in amplitude, but rather that they are not correlated with the dynamics of the macroscopic observables.

We feel that variational principle *II* is better justified, as there is no reason to suppose that random environmental noise should generally be small. Variational principle *II* requires us to make either the ergodic hypothesis for the bath random forces, or one must take the *definition* of a random force to mean that it is not correlated with macroscopic observables, either within a phase space average or in a time average. If experimentally we find that the random force is in fact correlated with a macroscopic observable, then we would know that there is an unidentified degree of freedom in the experimental system that is important to the dynamics of the identified macroscopic observables. Such a scenario can arise if there are neurons very far from the experimental electrodes that strongly drive the observed neurons, but that are too far from any electrode to be directly observed. In this case, there very well may be strong correlations between the macroscopic observables and the presumptive random force term. Thus, difficulty in minimizing the error functional $E_{II}(t_a, t_b)$ may sometimes signify that we have not identified all the key macroscopic observables. We may need to insert a few more electrodes in other parts of the brain.

In this case, if it is not possible to insert more electrodes into the brain or if we do not know where to insert them, what we can do is to take shorter time intervals of observation, given by $[t_a, t_b]$, and to extract the $K$'s and $G$'s for each time interval individually, a different set of $K$'s and $G$'s for each time interval. If the time intervals are short enough, then it will be possible to make $E_{II}(t_a, t_b)$ as small as one likes, simply because there will be fewer data points to be fitted with a given number of free fitting parameters. The $K$'s and $G$'s will then vary across time intervals. These variations reflect the influence of the unseen variables.

*3.1. Nonlinear interactions*

It may seem curious that Eq (7) is linear in the macroscopic variables $X(t)$ and $\dot{X}(t)$ even though we have made no assumptions about linearity in the microscopic variables, the $\bar{q}_i$'s and $\bar{p}_i$'s. The reason is a subtle one but important to understand. In Eq (2), every component of the vector $X(t)$, $x(n,t)$, is itself a vector in Hilbert space which can be expanded in terms of an infinite basis set of eigenfunctions $\phi_j$ of the Liouville operator $L$:

$$x(n,t) = \sum_{j=1}^{\infty} a_j(n,t)\, \phi_j(\Gamma) \qquad \text{Eq (31)}$$

Higher order functions, such as $x(n,t)^2$, $x(n,t)^3$, etc, have their own expansion coefficients in terms of these eigenfunctions, and the expansion coefficients will not necessarily have a simple relationship to those for $x(n,t)$. What this means is that in Hilbert space, higher order functions of the macroscopic variable $x(n,t)$ are considered additional dimensions in the abstract space. If they are important for describing the dynamics of $X(t)$, then one will not be able to satisfy one's variational principle and $E_{II}(t_a, t_b)$ will be in some sense too large. Difficulty in satisfying the variational principle thus can signal either that there are important unseen variables at play or that there are nonlinear dependences on the macroscopic variables. In the latter case, one remedy would be to define a vector $Z(t)$ with components $z_{N+k}(n,t) = x(n,t)^k$. One can include higher order terms up to whatever order $k$ one likes, in order to capture some of the nonlinear dependences. To generalize even further, one could begin including other functions as well, for example, sigmoidal functions for modeling nonlinear neuronal responses. Regardless of what one chooses, the vector $Z(t)$ will evolve according to the Liouville equation:

$$\frac{d}{dt} Z(t) \equiv \dot{Z}(t) = L Z(t) \qquad \text{Eq (32)}$$

The rest of the development follows as before, and one finds a corresponding equation of motion for $Z(t)$:

$$\ddot{Z}(t) = -K \circ Z(t) - G \circ \dot{Z}(t) + F_R(t) \qquad \text{Eq (33)}$$

One can still appeal to either variational principle $I$ or $II$ to extract the $K$'s and $G$'s from experiment data.

In general, there is no simple prescription for determining which macroscopic observables are important nor the best nonlinear forms for the interactions between the macroscopic observables. The choices that one makes will depend on the specifics of one's experimental setup (where one has inserted electrodes) and on one's intuition about how best to model explicit interactions. Nonetheless, the form of Eq (17) or Eq (33) constrains the mathematical model to one that is consistent with the laws of physics, and the variational principles *I* and *II* guarantee the best fit of one's model to experimental data.

As an example, consider the situation where there are two intracranial electrodes, and where the neurons near one electrode interact with neurons near the other through both a linear term and a sigmoidal term. There may also be a frictional force acting at each site. A reasonable set of equations of motion may then look like this:

$$\ddot{x}_1(t) = -K_{11} \circ x_1(t) - K_{12} \circ x_2(t) - K'_{12} \circ \tanh[\lambda x_2(t)]$$
$$- G_{11} \circ \dot{x}_1(t) + F_1(t)$$

Eq (34)

$$\ddot{x}_2(t) = -K_{21} \circ x_1(t) - K_{22} \circ x_2(t) - K'_{21} \circ \tanh[\lambda x_1(t)]$$
$$- G_{22} \circ \dot{x}_2(t) + F_2(t)$$

Eq (35)

One may then appeal to either variational principle to obtain the $K$'s and $G$'s, and even $\lambda$ if one wishes. The Zwanzig-Mori approach thus allows for a very flexible approach to the mathematical modeling of macroscopic observables. One does one's best to set up an equation of motion, with a certain set of fitting parameters, and then one appeals to the variational principle of one's choice to extract these fitting parameters from experiment. The variational principle guarantees the best description of the experimental data, given one's choice of a mathematical model.

*3.2. Piece-wise linear approximation*

Alternatively, one can ignore all macroscopic observables that are not linear functions of the $x(n,t)$'s and simply take sequential time intervals $[t_a, t_b]$ to be short enough that $E_I(t_a, t_b)$ or $E_{II}(t_a, t_b)$ is satisfactorily small. How short this time interval has to be is however short is necessary to obtain a value of $E_I(t_a, t_b)$ or $E_{II}(t_a, t_b)$ that is below one's desired threshold. In this case, one is assuming that the dynamics is linear within each time interval $[t_a, t_b]$, but not necessarily linear across time intervals. This approach is the piece-wise linear approximation. It is worthwhile considering when it is valid.

If the underlying microscopic dynamics obeys classical dynamics (i.e., Newtonian physics), then there is no loss of generality in making the piece-wise linear approximation, at least as far as the validity Eq (17) is concerned. The reason is that the force exerted on a classical particle is proportional to the slope of a potential energy function at a *single point* on that surface, with that point being given by the instantaneous coordinate of the particle $X(t)$. For a short enough period of time, one can perform a Taylor expansion of the potential energy surface about the instantaneous coordinate of the particle, expand to just the quadratic order term, and ignore the higher order terms. The force, since it is proportional to the slope of the potential energy surface, is then always piece-wise linear in $X(t)$.

In contrast, the instantaneous force on a quantum wavepacket depends on the shape of the potential energy surface over the instantaneous spatial extent of the entire wavepacket. If this wavepacket extends over a substantial patch of the potential energy surface, then an accurate Taylor expansion of the potential energy surface may have to include terms higher than just the quadratic order term. Thus quantum dynamics may not always be accurately described by a piece-wise linear assumption.

Fortunately, biological systems are typically far from the quantum regime, and in this case, we can always simply take the time intervals $[t_a, t_b]$ shorter and shorter until we satisfy our variational criterion. Piece-wise linear analysis (also known as instantaneous normal modes) has been surprisingly successful in describing even highly nonlinear dynamics, including that of liquids [16-20].

Once we have obtained the $K$'s and $G$'s, we can interpret interactions between groups of neurons with these functions. Which neuronal groups are linked by either the $K$'s or $G$'s? What are the time scales for the time delays? In particular, the eigenstates and eigenvalues of the $K$-matrix would be of high interest, as these eigenstates represent spatiotemporal patterns created by the interaction between the explicit degrees of freedom, i.e., these eigenstates may represent "memory traces".

To explore this idea, first ignore the convolutions in Eq (17) and consider the eigenstates of $K$ where the eigenvalues are purely real and positive. These eigenstates are oscillatory states which reside in free energy "wells". In terms of the EEG, one will see "standing wave" oscillations distributed over the spatial distribution of the respective eigenstates. These are stable states that can be used to store information. In contrast, eigenstates of $K$ where the eigenvalues are purely real and negative are unstable states which map onto free energy "barriers". In terms of the EEG, these states are evanescent states which do not recur, or at least not in a periodic way. These unstable states are not useful for storing information because they are transitory and one cannot reliably design a trajectory to return to such states.

Because $K$ is purely real but not necessarily symmetric, it can also have pairs of complex-valued eigenstates which have eigenvalues that are also complex. In each pair, the eigenvalues and eigenstates are complex conjugates of each other. The EEG dynamics represented by these pairs is that of a "traveling wave" that travels between the real and imaginary parts of the eigenstates. For instance, if one eigenstate is $\hat{e}_1 = \hat{e}_R + i\hat{e}_I$ with complex eigenvalue $\lambda_1 = \lambda_R + i\lambda_I$, then the other eigenstate is $\hat{e}_2 = \hat{e}_R - i\hat{e}_I$ with eigenvalue $\lambda_2 = \lambda_R - i\lambda_I$. In terms of the EEG dynamics, one will see activations of EEG voltages that morph continuously between the spatial distribution represented by $\hat{e}_R$ and that represented by $\hat{e}_I$. If $\lambda_R > 0$, then these states are stable and can also be used to store information.

The presence of the convolution between $K(\tau)$ and $X(t)$ in Eq (17) opens up the possibility of super-eigenstates that are not only spatially distributed, but also extended in time. To see how these arise, define a super-vector $\mathbf{X}(t)$ by concatenating $(M+1)$ consecutive $N$-dimensional $X$-vectors:

$$\mathbf{X}(t) = \begin{pmatrix} X(t) \\ X(t-\delta t) \\ \vdots \\ X(t-M\delta t) \end{pmatrix} \qquad \text{Eq (36)}$$

The number $M$ is determined by the number of time steps that contribute to the convolutions involving the $K$ and $G$ matrices. A super-vector $\mathbf{F}_R(t)$ can be defined in an analogous way, as well as $N(M+1) \times N(M+1)$ dimensional super-matrices $\mathbf{K}$ and $\mathbf{G}$. Equation (17) can then be written without convolution operators as:

$$\ddot{\mathbf{X}}(t) = -\mathbf{K} \cdot \mathbf{X}(t) - \mathbf{G} \cdot \dot{\mathbf{X}}(t) + \mathbf{F}_R(t) \qquad \text{Eq (37)}$$

The eigenstates of $\mathbf{K}$ now span not only space but also time over an interval given by $\tau_M = M\delta t$. We suggest that those eigenstates of $\mathbf{K}$ that are stable represent spatiotemporal memory traces.

*3.3. Criticality, neural and thermodynamic*

Of increasing interest in neurodynamics is the idea that the brain may poise itself near a critical point [21-23]. Near this neural critical point, neuronal elements may exhibit spatiotemporal patterns of correlated activation that span many length scales and time scales, from the length scales of just a few neurons to that encompassing macroscopic brain regions, and from time scales of individual action potentials (millisecs) to the much longer time scales of a completed thought (seconds to minutes, and possibly to even longer times). For neural systems, criticality can be defined in terms of the connectivity of the system [24, 25]. It has been shown that a neural system at critical connectivity exhibits maximal information storage capacity, and optimal information transmission and processing [21, 22, 24, 26-28]. Because the survival of an animal depends on how well its brain processes, stores and retrieves information, we have hypothesized that biological neural systems must maintain themselves at or near critical connectivity [29]. If this hypothesis is true, then failure to remain at or near critical connectivity may represent neurological disease, for instance, epilepsy [30]. If this hypothesis is false, it cannot be too far wrong, and it would be important to characterize how far a neural system can be from critical connectivity before it fails to be useful as an information processing system.

It would be of great interest to have a reliable measure of connectivity in the intact brain, as measured by clinically available electrodes either on the scalp of a human subject, placed on the surface of the brain, or inserted into the brain. For practical reasons, all such systems monitor only a tiny subset of all neural activity in the intact brain. An important consequence is the *subsampling problem*, where monitoring of only a portion of the total system is likely to misrepresent the true state of the system, including the possibility of mistaking a critical system for a subcritical or supercritical one [31]. It appears that one needs to monitor on the order of 25% of the total system in order to have some hope of a reliable measurement of system connectivity, at least using current measures of connectivity (Priesemann V; Personal communication, 2009).

Can the projection operator approach afford a more reliable measure of criticality, and can it connect the neural concept of criticality to the more traditional thermodynamic concept [32]? We do not know as yet, but the conceptual framework of the projection operator approach is suggestive of an approach to this problem. The reasoning is as follows.

The vector of macroscopic observables, $X(t)$, can be regarded as a generalized coordinate of an abstract neural "super-particle" that exists in a very high-dimensional space. The coordinate reflects the collective movement of a macroscopic number of microscopic charged ions and molecules in a very complicated, nonlinear way, as given by Eq (1). Notwithstanding the complexity of the microscopic dynamics, it is a characteristic of critical phenomena to span all length and time scales [32], and thus one may ask, if we define a heat capacity for the macroscopic neural super-particle, will it behave in the same way as the heat capacity of the microscopic system if either system is at or near its critical point? The answer is yes, it should, and therefore the critical point of the microscopic system should be identical to that of the macroscopic system. The heat capacity of the neural system should diverge at the critical point in the same way, *with the same critical exponent*.

To explore this idea, note that changes in the internal energy $U(t)$ of the oscillator are driven by the force $-K \circ X(t)$, and so incremental changes in this energy are given by

$$\delta U(t) = -\delta X \cdot [K \circ X(t)] \qquad \text{Eq (38)}$$

If we take the kinetic energy to be $K_E(t) = \frac{1}{2} \dot{X}(t) \cdot \dot{X}(t)$ and the temperature to be related to the kinetic energy through

$$K_E(t) = \frac{N}{2} k_B T(t) \qquad \text{Eq (39)}$$

where $T(t)$ is the instantaneous temperature, $N$ is the total number of EEG channels and $k_B$ is Boltzmann's constant [33], we can then define a dimensionless internal heat capacity as

$$C(t) = \frac{\delta U(t)/\delta t}{\delta K_E(t)/\delta t} = -\frac{\dot{X}(t) \cdot K \circ X(t)}{\dot{X}(t) \cdot \ddot{X}(t)} \qquad \text{Eq (40)}$$

Using Eq (40), we may then investigate divergences of the heat capacity near phase transitions including the critical point. Other thermodynamic quantities of interest may be similarly defined [34]. A caution is that more sophisticated definitions of temperature than Eq (39) may be needed for systems that are not at thermal equilibrium.

One can also begin to think of EEG dynamics in terms of established particle dynamics concepts. What does the free energy surface $U(t)$ look like on which brain dynamics moves? Are there deep energy wells that trap neural super-particle trajectories? Are there frequent hops between energy wells? What proportion of time is spent inside a well vs on an energy barrier? How high are the energy barriers to transitions between free energy wells? Can one hope to bring into play such landmarks of physical chemistry as transition state theory and other reaction rate theories? What happens in disease, for example, in epilepsy? Do the energy wells become shallower? Is there less friction? We have previously taken small steps in investigating these questions [35], but a great deal more is possible. Our prior experience shows that the calculations required to utilize variational principle *II* are feasible.

*3.4. Causality*

According to Newtonian physics, the force exerted by one body on another acts instantaneously, with no time delay. Projection operator theory shows us that if the force exerted by one macroscopic body on another is mediated through many microscopic degrees of freedom, the response of the second body may be delayed in time. This time delay is represented by the convolution in Eq (17). The macroscopic time delay is intuitively obvious from our daily experience, and does not surprise us. The time delay is suggestive of but not proof of causality. Can causality be demonstrated using the Zwanzig-Mori equation of Eq (17)?

In this regard, it is instructive to consider Granger causality [36], which is finding increasing application in neuroscience and many other areas. In Granger causality, one assumes that the value of $X(t+\delta t)$ at time $t+\delta t$ depends on all prior values at earlier times. One then asks if $X(t+\delta t)$ also depends on prior values of another vector $Y(t)$. To answer this question, one can construct two time series:

$$X(t+\delta t) = G \circ X(t) + R_X(t) \qquad \text{Eq (41)}$$

$$X(t+\delta t) = H_1 \circ X(t) + H_2 \circ Y(t) + R_{XY}(t) \qquad \text{Eq (42)}$$

One solves for *G* in Eq (41) by minimizing an error functional over some time interval $[t_a, t_b]$:

$$E_X(t_a, t_b) = \int_{t_a}^{t_b} dt \left[ R_X(t) \cdot R_X(t) \right] \qquad \text{Eq (43)}$$

Similarly one solves for $H_1$ and $H_2$ by minimizing the error functional

$$E_{XY}(t_a, t_b) = \int_{t_a}^{t_b} dt \, [R_{XY}(t) \cdot R_{XY}(t)] \quad \text{Eq (44)}$$

If $E_{XY}(t_a, t_b)$ is significantly less than $E_X(t_a, t_b)$ by some statistical measure, then one says that $Y(t)$ Granger causes $X(t)$.

The Zwanzig-Mori formulation can be applied to demonstrate causality in an analogous way. The Granger error terms $R_X(t)$ and $R_{XY}(t)$ correspond to random force fluctuations in the Zwanzig-Mori equation. The error functionals of Granger causality in Eqs (43) and (44) are equivalent to the Zwanzig-Mori error functional $E_I(t_a, t_b)$, corresponding to our variational principle *I*. That the Zwanzig-Mori equation contains both a coordinate and a velocity term is not a major difference, as within the Granger formulation, one could simply take the velocity as another degree of freedom, in effect, another coordinate.

However, within the Zwanzig-Mori formalism, one also has the option of appealing to variational principle *II*. This variational principle may be more useful for noisy systems. For noisy systems, it is not justifiable to assume that the random force contribution to dynamics should be as small as possible. Random environmental forces may in fact be dominant. Nonetheless, one can still define the random force to have no time correlation with the macroscopic observables, for reasons discussed above. Thus, we feel that variational principle *II* has advantages over variational principle *I*.

*3.5. Prediction and control: relation to Kalman filters*

Kalman filters are also beginning to find application in engineering applications of neural prediction and control [37]. The idea here is to understand how a system responds to externally applied perturbations so that one can apply the perturbations in a rationally planned way so as to achieve a desired response.

Zwanzig-Mori theory also allows one to probe system response to external perturbations. Let the system observables be described by a vector $X(t)$ and let the external perturbations be described by a time-dependent vector $Y(t)$. On the experimental system, one may apply a variety of representative test perturbations $Y(t)$. One then constructs a vector $Z(t)$ containing both vectors $X(t)$ and $Y(t)$:

$$Z(t) = \begin{bmatrix} X(t) \\ Y(t) \end{bmatrix} \qquad \text{Eq (45)}$$

Going through all the steps as before, we find the Zwanzig-Mori equation for $Z(t)$

$$\ddot{Z}(t) = -K \circ Z(t) - G \circ \dot{Z}(t) + F_R(t) \qquad \text{Eq (46)}$$

Appealing to either variational principle *I* or *II*, we can extract all the $K$'s or $G$'s in Eq (46) as a result of the test perturbations. There will be off-diagonal elements of the $K$'s and $G$'s that describe the effect of a given test perturbation $Y(t)$ on the future dynamics of $X(t)$. Knowing these elements allows one to predict the future response of the system to any variation of the test perturbations.

*3.6. Related approaches*

Hänggi and coworkers have also taken advantage of the Zwanzig-Mori formulation of the GLE to construct a hierarchy of statistical measures of system memory [38]. These measures have been applied to a patient with photosensitive epilepsy with striking results [39]. The Zwanzig-Mori approach has also been applied to construct a renormalized kinetic theory of dense fluids [40].

In terms of other theoretical approaches for deriving macroscopic equations of motion based on microscopic dynamics, a moment expansion method is also possible. In this approach, one defines moments of the macroscopic observables, e.g., $\langle X(t)^n \dot{X}(t)^m \rangle$ where *m* and *n* are positive integers and where the angular brackets signify a phase space average over a distribution function, which need not be an equilibrium distribution function. Taking time derivatives and applying microscopic laws within the angular brackets results in coupled differential equations linking the dynamics of the different order moments of $X(t)$. In general, the lower order moments will depend on higher order moments, and one needs a way of closing the moment expansion [e.g., see Refs 41, 42].

Yet another method is to expand the distribution function in terms of a "basis set", and then derive equations of motion for the expansion coefficients. It is possible to allow the basis functions themselves to change in time by appealing to the Dirac-Frenkel variational principle [43], which resembles our variational principle *II*. Indeed, the Dirac-Frenkel variational principle was the motivation for variational principle *II*. The variationally optimized "mobile basis set" approach has been applied to quantum dynamics [44, 45] and can also be applied to statistical mechanics. A suggestion for workers who wish to try this mobile basis set approach is that one should expand the square root of the distribution function in a basis set, not the distribution itself, or else one would not be able to maintain normalization simultaneously with energy conservation.

*3.7. Summary*

The Zwanzig-Mori formalism is a simple and flexible mathematical framework for interpreting macroscopic dynamics even in the presence of significant environmental noise. Its range of applicability is very broad, including the dynamics of all natural systems governed by classical or quantum physics. It is based on sound physical principles, and it allows one to extract model parameters from experimental data using one of two variational principles. These variational principles are our principal contribution to the formalism.

**Acknowledgements**
DH was supported by NIH 1KL2RR025012-01 and by the NIH Loan Repayment Program. We thank Drs. John Straub, Tom Keyes and Viola Priesemann for helpful comments.